# Resource Allocation in a Network-Based Cloud Computing Environment: Design Challenges

M. Abu Sharkh, M. Jammal, A. Shami*, and A. Ouda

*Abstract*—Cloud computing is an increasingly popular computing paradigm, now proving a necessity for utility computing services. Each provider offers a unique service portfolio with a range of resource configurations. Resource provisioning for cloud services in a comprehensive way is crucial to any resource allocation model. Any model should consider both computational resources and network resources to accurately represent and serve practical needs. Another aspect that should be considered while provisioning resources is energy consumption. This aspect is getting more attention from industry and governments parties. Calls of support for the green clouds are gaining momentum. With that in mind, resource allocation algorithms aim to accomplish the task of scheduling virtual machines on data center servers and then scheduling connection requests on the network paths available while complying with the problem constraints. Several external and internal factors that affect the performance of resource allocation models are introduced in this paper. These factors are discussed in detail and research gaps are pointed out. Design challenges are discussed with the aim of providing a reference to be used when designing a comprehensive energy aware resource allocation model for cloud computing data centers.

## I. INTRODUCTION

CLOUD computing is an increasingly popular computing paradigm, now proving a necessity for utility computing services. Several providers have Cloud Computing (CC) solutions available, where a pool of virtualized and dynamically scalable computing power, storage, platforms, and services are delivered on demand to clients over the Internet in a pay as you go manner. This is implemented using large Data Centers (DCs) where thousands of servers reside. Clients have the choice between using private clouds which are DCs specialized for the internal needs of a certain business organization and public clouds which are open over the Internet to the public for use. Services are offered under several deployment models, Infrastructure as a Service (IaaS), Platform as a Service (PaaS), Software as a Service (SaaS) and Network as a Service (NaaS). Each provider offers a unique service portfolio with a range of options that include Virtual Machines (VMs) instance configuration, nature of network services, degree of control over the rented machine, supporting software/hardware security services, additional storage, etc. To move to the cloud, clients demand guarantees with regards to achieving the required improvements in scale, cost control, and reliability of operations. Despite its importance, providing computation power alone is not sufficient as a competitive advantage. Other factors have gained more weight recently such as the networking solution offerings. The network performance and resource availability could be the tightest bottleneck for any cloud. This is seen as an opportunity for network service providers who are building their own clouds using distributed cloud architecture.

Here, we see the need for a comprehensive Resource Allocation (RA) and scheduling system for CC Data Center Networks (DCN). This system would handle all the resources in the cloud providers' DCN and would manage client requests, dictate RA, ensure satisfaction of network QoS conditions, and eliminate performance hiccups while minimizing the service provider cost and controlling the level of consumed energy.

The resource management of the DCs' servers and the network resources while scheduling and serving thousands of client requests on Virtual Machines (VMs) residing on DC servers, is a critical success factor. First, it is a main revenue source to the service provider as excess resources translate directly to revenue. Second, it is a key point that will make or break potential clients' decision to move fully to the cloud.

The previous RA models can be classified into three categories:

*A- Efforts with a focus on DC Processing Resources:*
Multiple models were proposed previously, where resources are scheduled based on user requests. In [1], a queuing model is proposed where a client requests virtual machines for a predefined duration. Jobs are assumed not to communicate with each other or transmit or receive data. No preference is required as to where the VMs are scheduled. In [2], an algorithm is proposed to optimally distribute VMs in order to minimize the distance between them in a DC grid. The only network constraint used is the Euclidean distance between DCs. No specific connection requests or user differentiation is used. In the same paper, an algorithm is proposed to schedule VMs on racks, blades and processors within one DC to minimize communication cost.

*B- Efforts with a focus on DC Network Resources:*
In [3], the authors tackle the problem where a client may have multiple jobs being processed at the same time but not necessarily on the same server. Requests are abstracted as a Virtual Network (VN) where VMs represent nodes and paths between two nodes represent VN links. The problem is treated as an optimization problem of provisioning a virtual network with the objective of revenue maximization. It did not introduce reservation start time or duration. The scenario where a user wants to request more connectivity for an already reserved VM is not considered. In [4], authors tackle the problem of proposing the best virtual network with IP over Wavelength Division Multiplexing (WDM) network. Constraints are based on propagation delay, flow conversion constant and capacity.

*Corresponding Author

*C- Efforts with a focus on Energy efficient DC RA:*

Multiple solutions were proposed with the aim of reaching an energy efficient RA scheme. A common concept is the idea used in [5] which is to consolidate tasks or VMs on the least number of servers and then switch the unused servers off or make them idle. The problem is modeled as bin-packing problem with the assumption that servers are the bins and they are full when their resources reach a predefined optimal utilization level. Power consumption by network components is not considered.

Other works took a hardware planning approach to the problem. Instead of targeting the highest performance possible, they aim at executing a certain work load with as little energy as possible. This would not suit the cloud clients' needs as this architecture does not support applications with high computational demands.

An economic approach to manage shared resources and minimize the energy consumption in hosting centers is described in [6]. The authors present a solution that dynamically resizes the active servers and responds to the thermal or power supply events by downgrading the service based on the Service Level Agreement (SLA). With the scheduling component already allocating the requests at the lower limit of the SLAs to have enough resources, it will not be easy to find requests that can tolerate downgrades.

## II. NETWORK AWARE RA: DESIGN CHALLENGES

*A- A Comprehensive Solution for Network Processing RA*

Provisioning for cloud services in a comprehensive way is crucial to any RA model. Any model should consider both computational resources and network resources to accurately represent practical needs. First, excluding the computational resources during the design of the RA model deprives the model of the main cloud service. Cloud DCs are built first and foremost as ways to outsource computational tasks. Any model that optimizes DC resources should include answers to questions like: How are VMS allocated? How are processing resources modeled? What is the resource portfolio that is being promoted to clients? How the DC resources are distributed physically? The other side of the coin is networking services. As clients ask for tasks to be processed in the DC, they need networking service with adequate QoS standards to send and receive their application data.

As reported in [7], only 54% of the IT professionals surveyed about their use of cloud services indicated that they involve network operations' personnel, down from 62% in 2009. This directly affects the use of network best practices and the attention to the health of overall traffic delivery. Also in [7], 28% of survey respondents believed that monitoring and troubleshooting packet traces between VMs is required. In addition, 32% believed that monitoring and troubleshooting traffic data from virtual switches is required.

Bandwidth costs deeply affect the cloud clients' financials. Microsoft azure for example charges clients for download based on the exact amount downloaded. Downloading around 950 GB/month costs the client $113/month. In comparison, Comcast -the largest Internet provider in the US- offers a plan with a bandwidth that can download the same amount at $40/month. Azure offers free upload and free data exchange between VMs that are located in the same datacenter. However, the price difference is an issue clients will consider. Therefore, optimizing the bandwidth cost represents an opportunity of profit for providers and an opportunity of saving for clients.

The network resources weight in the cloud market has alerted network service providers to build their own distributed DCs with a vision to enter the CC market. They envision replacing a large DC with multiple smaller DCs to be closer to the clients. This setup turns the network infrastructure into a distributed cloud. That in turn helps in controlling costs and increasing service differentiation.

A cloud service provider caters network services to clients to support one of three functions [8]:

1- Connecting the clients' private cloud (or headquarters) to VMs the client reserved in the DCs; using Internet or VPNs as shown in Figure 1.
2- Connecting the VMs on different public clouds to facilitate data exchange between two VMs reserved by the same client.
3- Connecting VMs on the same public cloud together

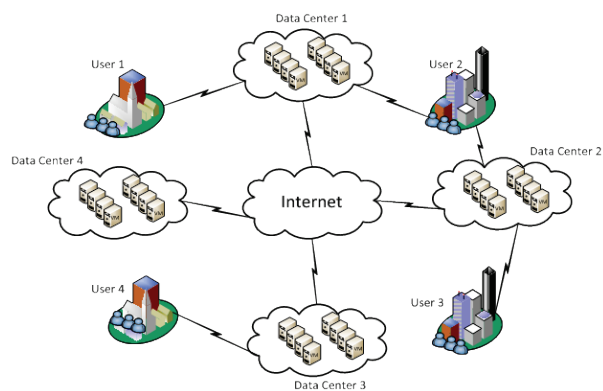

Figure 1: A sample network of private and public clouds connected through Internet or VPNs

It is no use to the clients if their application is producing the results needed in the required time if these results cannot be delivered to them through a stable network connection. In [9], data transfer bottlenecks are stated as one of the main obstacles cloud client growth is facing. The authors show that when moving large amounts of data in a distributed DC environment, the network service performance will be a critical point for the whole process. In the example mentioned, the authors reached the conclusion that the data transmission tardiness can cause the client to prefer sending data disks with a courier (FedEx, for example).

*B- Main Design Challenges:*

Targeting a network-aware RA system brings to the front multiple challenges that face the CC community. Addressing those issues would be of utmost importance to form a complete solution. These design challenges can be classified into external challenges which are enforced by factors outside the RA process (illustrated in Figure 2) and internal challenges that are related to the RA algorithm (shown in Figure 3).

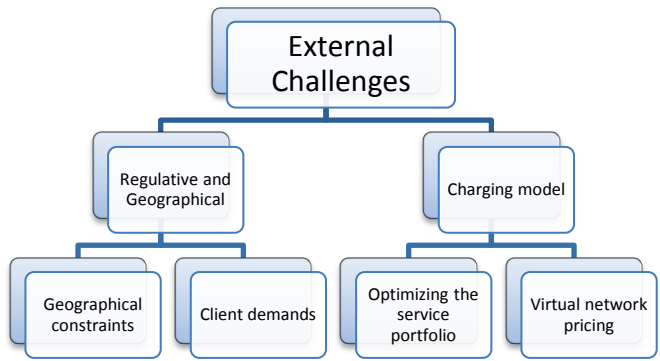

Figure 2: External challenges

*External Challenges*
1- Regulative and Geographical Challenges:

In the virtualization model used in cloud offerings, the client does not manage the physical location of data. Also, there is no guarantee given by the provider as for the data physical location in a certain moment [9]. In fact, it is a common practice to distribute client data over multiple geographically distant DCs. Splitting the data will enhance fault tolerance, but it presents regulative and security challenges. An example would be the regulative obligation of complying with the U.S. Health Information Portability and Accountability Act (HIPAA) (the Health Information Protection Act (HIPA) in Canada). HIPAA does not apply directly to third party service providers, it is imperative that health care organizations require the third-party providers to sign contracts which require them to handle all patient data in adherence with HIPAA standards. This raises some constraints to handling and storing data:

a- Geographical constraints: HIPAA requires that patient data does not leave US soil. This constraint limits the choice of DCs to allocate a VM to and limits data movement maneuvers while trying to optimize performance. Additionally, when data is stored in the cloud, it is necessary to know the physical location of the data, the number of data copies, data modification details, or data deletion details.

b- Client actions: To get more assurance about data security, clients may require guarantees like instant data wiping (writing over byte by byte) instead of deletion. They might also require storing encrypted data on the cloud. This would pose extra pressure on the performance and will make it harder to comply with QoS requirements.

c- Under HIPPA, patients have the right to access any information stored about them. A careful study of the locations of the patients and the usage distribution of these patients is crucial for the RA system. Considering this factor when placing the data would minimize the distance patient data will travel in the network. Making a decision where the data is located has a direct effect on minimizing the cost.

2- Charging Model Issues:

The resources management system should incorporate the clients charging model. For example, when using Amazon EC2, a client can pay for the instances completely on demand, reserve an instance for term contract or choose spot instances that enable him to bid for unused Amazon EC2 capacity. Issues to be considered here include:

A- Finding the service portfolio offering that maximizes the revenue weight of excess resources in the DC. Examining the options available in the market, it is clear that cost is not calculated based on static consumptions.

B- Finding the best way to integrate the virtual network usage into the cost analysis. Challenges would arise because a virtual link length/distance (and in turn cost) varies from link to link. A virtual link could even change to use another physical path on the substrate network based on the methodology used.

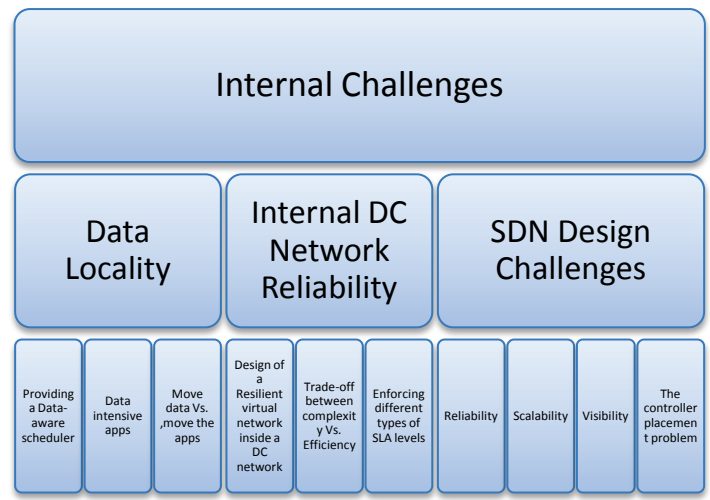

Figure 3: Internal challenges

*Internal challenges*
1- Data Locality: Combining Compute and Data Management

There is a need for systems to implement data locality features "the right way". This means how to combine the management of compute (processing) and data (network) resources using data locality features to minimize the amount of data movement and in turn improve application performance/scalability while meeting end users security concerns. It is important to schedule computational tasks close to the data, and to understand the cost of moving the work as opposed to moving the data.

To have a full view of how to use data locality these issues need to be considered:

a- A data aware-scheduler is critical in achieving good scalability and performance. A more specific perspective needs to be reached. This includes answering questions like: How much would the scheduler know at a certain moment? What are the policies and decision criteria for moving data? What data integration policies should be enforced?

b- Analyzing the behavior of data intensive applications is a good starting point to understand data locality and data movement patterns.

c- Also an idea to be evaluated is moving the application itself to servers in the DC where the needed data is. This raises questions about the availability of servers in the other DC, policy/ algorithm specifications regarding when to move the application considering that future demand might need data sets that are stored in the original location, decision criteria regarding whether to migrate the whole VM or just move the concerned application.

2- Reliability of Network Resources inside a DC

The DC internal network affects the performance deeply. The DC internal network design decisions affect performance and reliability of the DC resources. These decisions relate to factors like network topology, traffic routing, flow optimization, bandwidth allocation policies and network virtualization options.

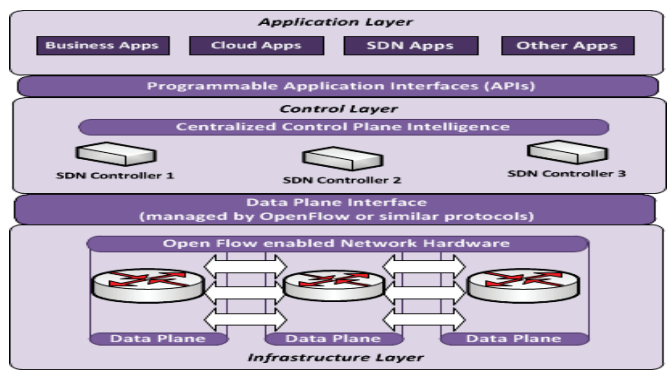

Figure 4: SDN architecture [10]

3- SDN Design Challenges inside the DCs:

SDN is a networking paradigm in which the forwarding behavior of a network element is determined by a software control plane decoupled from the data plane. This paradigm can enable many advantages if it was coupled with an efficient RA model. SDN leads to many benefits such as increasing network and service customizability, supporting improved operations and increased performance. The software control plane can be implemented using a central network controller which can handle the task of RA in the DCN by directing all the client requests to it. This controller will execute the RA algorithms then send the allocation commands across the network. Figure 4 shows a view of the SDN architecture.

Since it is a relatively new paradigm, the community still has to tackle deeply these issues regarding SDN:

A-Reliability - Using centralized SDN controller affects reliability. Although solutions like stand by controllers or using multiple controllers for the network are suggested, practical investigation is needed to reveal the problems and analyze the trade-offs of using such solutions.

B- Scalability - When the network scales up in the number of switches and the number of end hosts, the SDN controller becomes a key bottleneck. For example, [11] estimates that a large DC consisting of 2 million virtual machines may generate 20 million flows per second. The current controllers can support about $10^5$ flows per second in the optimal case [12]. Extensive scalability results in losing visibility of the network traffic, making troubleshooting nearly impossible.

C- Visibility - Prior to SDN, a network team could quickly spot, for example, that a backup was slowing the network. The solution would then be to simply reschedule it. Unfortunately with SDN, only a tunnel source and a tunnel endpoint with User Datagram Protocol (UDP) traffic are visible. One cannot see who is using the tunnel. Determining the origin of the problem is a challenge. The true top talker is shielded from view by the UDP tunnels, which means that when traffic slows and users complain, pinpointing the problem area in the network is a challenge. With the loss of visibility, troubleshooting is hindered, scalability is decreased and a delay in resolution could be quite detrimental to the business.

D- The controller's placement problem influences every aspect of a decoupled control plane, from state distribution options to fault tolerance to performance metrics. This problem includes placement of controllers with respect to the available topology in the network and the number of needed controllers. The placement is related to certain metrics defined by clients like latency, increasing number of nodes, etc. According to [13], random placement for a small value of *k* medians will result in an average latency between *1.4x* and *1.7x* larger than that of the optimal placement.

Thus cloud clients will see network service specification as a decisive factor in their choice to move to the cloud or to choose their cloud provider. Factors like bandwidth options, port speed, number of IP addresses, load balancing options and availability of VPN access should be considered by any comprehensive model.

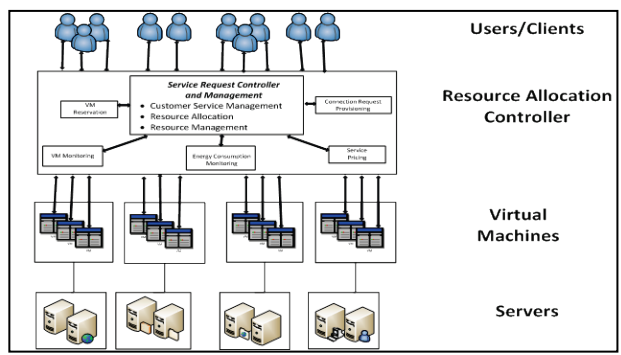

Figure 5: Cloud computing architecture

4- Fault Tolerance vs. Performance:

Despite its several applications and wide range acceptance, the current Cloud computing (CC) technology is still prone to hardware, VM and application failures. Therefore, a stable and efficient Fault Tolerance (FT) strategy is a crucial requirement to achieve availability, security and reliability of CC services and real time applications as well as ensuring seamless task execution.

Due to the complexity and inter-dependability of FT, implementing it in CC requires delicate analysis and consideration. CC requires autonomic FT policies for instances of VM applications. These techniques must integrate with workflow scheduling algorithms and synchronize among different clouds. Furthermore, CC requires either reactive FT or proactive FT based on the application type and level [14]. Reactive FT techniques, such as Restart, Replay and Retry, reduce the faults' effect on the application execution. On the other hand, the proactive FT techniques, such as Software Rejuvenation and Preemptive migration, predict faults and errors and get rid of the paralyzed components. Hence, in the context of FT, CC providers aim to implement failure recovery, cost-aware and performance effective FT policies.

FT strategy affects how VMs are distributed across the fault domains. This distribution often contradicts performance. The challenge here is to find the fault domain definitions and VM distribution that complies with fault tolerance constraints without compromising the performance.

5- Portability and Vendor Lock in:

This issue is a concern for the cloud clients. Clients require guarantees of the applications being portable and easily movable to other cloud providers. This affects VM deployment design and raises a concern for cloud providers regarding the optimal procedure when a certain client leaves. Which RA adjustments are made and how? Here, designing an efficient procedure is a big performance booster. Figure 5 shows where the RA controller lies in the cloud computing architecture. The figure summarizes the main RA functionalities in a cloud computing data center that are performed by the multiple modules of the RA controller.

III. ENERGY EFFICIENT NETWORK BASED RA

As DCs number and average size expand, so does the energy consumption. Electricity used by servers doubled between 2000 and 2005, from 12 to 23 billion kilowatt hours [15]. This is not only due to the increasing amount of servers per DC, but also the individual server consumption of energy has increased too. The increase in energy consumption is of major concern to the data center owners because of its effect on the operational cost. It is also a major concern of governments because of the increase in DCs' carbon footprint. The cloud client base is expanding by the day. This demand will lead to building new DCs and developing the current ones to include more servers and upgrade the existing servers to have more functionality and use more power. Power-related costs are estimated to represent approximately 50% of the DC operational cost, and they are growing faster than other hardware costs [16]. Thus, energy consumption is a major obstacle that would limit the providers' ability to expand. Recently, the response to this fact is seen in the practical landscape as major players in the cloud market are taking more serious steps. Companies as large as Microsoft and Google are aiming to deploy new DCs near cheap power sources to mitigate energy costs [16]. Recently, leading computing service providers have formed a global consortium, the Green Grid, which aims at tackling this challenge by advancing energy efficiency in DCs. This is also pushed by governments in an attempt to decrease the carbon footprints and the effect on climate change. For example, the Japan Data Center Council has been established to mitigate the soaring energy consumption of DCs in Japan.

A- A comprehensive Solution for Energy Efficient Network-based RA

Any model that aims at allocating resources while minimizing energy consumption in a distributed cloud should consider all sources of energy consumption. It should include analysis for power used by CPU, memory, hard disks, and power supply unit in a server. An illustration of the power consumption of the possible server components is shown in Figure 6.

Also the model should investigate power consumed by network components to transmit data inside and outside the DC. Although the power consumed by a cable or a router for example is a small percentage of a power consumed by a server rack, the large number of devices that local and global networks consist of consumes significant amounts of power. Hardware design optimization is a direction researchers aim at when trying to minimize power consumption. However, the most rewarding concept to save power is to optimize the network performance. Moving data using shorter paths and flow optimization cause significant savings. An efficient VM placement technique affects directly the number of network components used per connection. An efficient data-aware scheduler can be the difference between moving data within the same rack, using the local network within a data center or sending the data to another one across the ocean. Any energy gain from any of these methods is an important achievement since one DC's operational cost impact on environment is high.

B-  Common solutions and common trade-offs

1-A solution with multiple variations in the literature is consolidation of applications on fewer servers. This concept, despite its positive effect on power consumption, affects the performance negatively. There are three main issues here:

  a- A consolidation could quickly cause I/O bottlenecks. Concentration of VMs increases the competition for physical server resources which threatens the performance as it has a high probability of having I/O bottlenecks. In addition to that, it can cause more power consumption because of the latency in task completion.

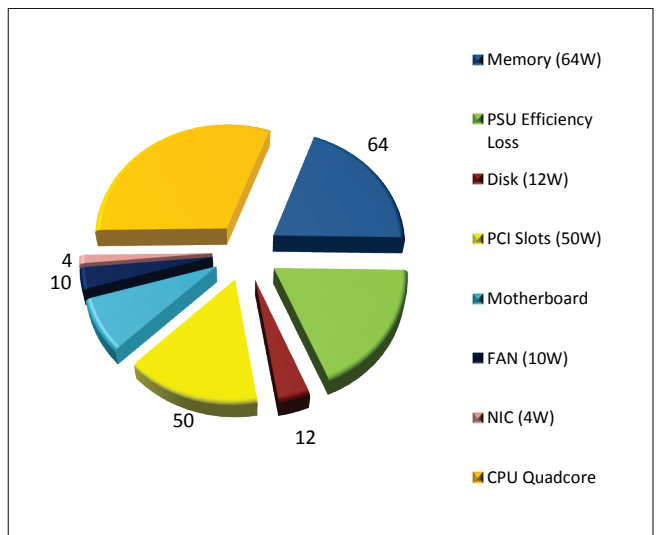

Figure 6: Server Power Consumption [14]

  b- Network bottlenecks: Connection blocking would increase visibly as connections from and to all the consolidated VMs compete for the links available to the physical node holding the server. For applications with heavy data transaction, higher blocking percentage would be found around the servers carrying the consolidated VMs. This would cause even more latency and would consume more network related power.

  c- The method used to hibernate or shut down unused servers should be considered. There is the latency and power consumption caused by the system hibernation and waking up. If used, consolidation should be part of a more global solution that takes in consideration those issues along with client priorities. In [17], the authors explain the energy waste that happens because of idle servers. "Even at a very low load, the power consumed is over 50% of the peak power." This is more apparent when there is a bottleneck since all the other idle resources are wasting power.

  2- VM migration is the core of the consolidation process. The methodology might differ based on the VM size and configuration variations. Nevertheless, trade-offs have to be considered between the power gained by moving the VM and hibernating the machine it is on and the total losses caused by this migration. These losses include:

  a- Time lost moving VM through the network.
  b- Power consumed by network components during the move
  c- Latency of the task completion caused by the changed node on the network and the need to provision new network resources.

C-  Energy Consumption vs. Optimal Performance: Hardware Contradictions

  The way processors work currently, a higher performance is achieved by maximizing the use of the processor cache memory and minimizing the use of the main memory and disks. In addition, using mechanisms like out-of-order execution, high speed buses and support for a large number of pending memory requests increases the transistor counts which lead to more wasted power. Thus, the question of the optimal point between performance and power consumption will arise.

D- Cooling Challenges

  A considerable amount of the electrical energy consumed by the computing and network resources is transformed into thermal energy. This thermal energy reduces the data centers devices lifetime and affects the system availability negatively. Therefore, dissipation of such energy is a crucial requirement in any cloud infrastructure in order to protect devices from failure, crashes and maintain them at safe operating point. As reported in [18], the initial cost of buying and installing the infrastructure of a data center with 1000 computing racks is between $2-$5 million. However, the cooling system costs annually around $4-$8 million. For this reason, software-side optimization might be a promising solution to mitigate the cooling system problem.

IV. CONCLUSION

Nowadays, CC shows its paramount importance for computing services. To reach a complete RA solution for managing CC DCs, optimizing computational resources, network resources and energy consumption are the main sides. This paper introduces some internal and external factors that affect the design of DC RA models. External challenges are mostly caused by regulative, geographical and charging model related factors. Internal challenges include maximizing the benefits from data locality features. They also include designing a reliable internal DC network. Other internal factors are related to SDN, fault tolerance and portability. Designing an energy-aware RA model faces performance challenges that are mostly caused by consolidation, VM migration and server idle state configuration. These design challenges are discussed with the aim of providing a reference to be used when designing a comprehensive energy aware resource allocation model for CC data centers.

REFERENCES

[1] S. Maguluri, R. Srikant, and L. Ying, "Stochastic Models of Load Balancing and Scheduling in Cloud Computing Clusters", IEEE INFOCOM 2012 Proceedings. pp.702 710, 25-30 Mar, 2012.

[2] M. Alicherry and T.V. Lakshman, "Network Aware Resource Allocation in Distributed Clouds", IEEE INFOCOM 2012 Proceedings, pp.963-971, 25-30 Mar, 2012.

[3] G. Sun, V. Anand, H. Yu, D. Liao, and L.M Li, "Optimal Provisioning for Elastic Service Oriented Virtual Network Request in Cloud Computing", IEEE Globecom 2012, pp.2541-2546, 3-7 Dec. 2012

[4] B. Kantarci, and H.T. Mouftah, "Scheduling Advance Reservation Requests for Wavelength Division Multiplexed Networks with Static Traffic Demands", IEEE Symposium on Computers and Communications (ISCC), pp. 806-811, 1-4 Jul 2012.

[5] S. Srikantaiah, A. Kansal, and F. Zhao, "Energy Aware Consolidation for Cloud Computing", Cluster Computing, vol. 12, pp. 1-15, 2009.

[6] J. S. Chase, D. C. Anderson, P. N. Thakar, A. M. Vahdat and R. P. Doyle, " Managing Energy and Server Resources in Hosting Centers", Presented at 18th ACM Symposium on Operating Systems Principles (SOSP'01), Oct 21, 2001.

[7] J. Frey, "Network Management and the Responsible, Virtualized Cloud", Research Report, Feb 2011, online report available at http://www.enterprisemanagement.com/research/asset.php/1927/Network-Management-and-the-Responsible,-Virtualized-Cloud.

[8] M. Abu Sharkh, A. Ouda and A. Shami, "A Resource Scheduling Model for Cloud Computing Data Centers", The 9th International Wireless Communications & Mobile Computing Conference (IWCMC 2013) July 1-5, 2013 – Cagliari, Sardinia – Italy.

[9] M. Armbrust, A. Fox, R. Gri_th, A. Joseph, R. Katz, A. Konwinski, G. Lee, D. Patterson, A.Rabkin, I. Stoica, and M. Zaharia, "Above the Clouds: A Berkeley View of Cloud Computing", Tech. Rep. UCB/EECS-2009-28, EECS Department, U.C. Berkeley, Feb 2009.

[10] M. Jammal, Ta. Singh, A. Shami, R. Asal, and Yi. Li, " Software Defined Networking: A Survey," (In Submission) IEEE Communications Surveys and Tutorials, IEEE, July 2013

[11] A. Tavakoli, M. Casado, et al., "Applying NOX to the Datacenter", In Proc. HotNets (Oct 2009).

[12] Pluribus Networks, "Of Controllers and Why Nicira Had to Do a Deal, Part III: SDN and Openflow Enabling Network Virtualization in the Cloud)", August 2012, see http://pluribusnetworks.com/blog/

[13] B. Heller, R. Sherwood and .,N. McKeown, "The Controller Placement Problem", In Proceedings of the First Workshop on Hot Topics in Software-defined Networks (HotSDN 2012), pages 7–12, Aug. 2012.

[14] A. Tchana, L. Broto, D. Hagimont,, "Approaches to cloud computing fault tolerance," Computer, Information and Telecommunication Systems (CITS), 2012 International Conference on , pp.1,6, 14-16 May 2012

[15] L. Minas and B. Ellison, "The Problem of Power Consumption in Servers", Prepared in Intel Lab, Dr. Dobbs Journal, Mar 2009.

[16] B.G. Chun, G. Iannaccone, G. Iannaccone, R. Katz, G. Lee, L. Niccolini, "An Energy Case for Hybrid Datacenters", ACM SIGOPS Operating Systems Review, v.44 n.1, Jan. 2010

[17] G Chen et al., "Energy-Aware Server Provisioning and Load Dispatching for Connection-Intensive Internet Services", Proceedings of the 5th USENIX Symposium on Networked Systems Design and Implementation, p.337-350, Apr 16-18, 2008.

[18] C.D. Patel, C.E. Bash, and A.H. Beitelmal, "Smart Cooling of Data Centers," Hewlett-Packard Development Company L.P. Patents, 2003